\documentclass[letterpaper,10pt,conference]{ieeeconf}

\IEEEoverridecommandlockouts
\usepackage[utf8]{inputenc}

\newcommand\oprocendsymbol{\hbox{$\square$}}
\newcommand\oprocend{\relax\ifmmode\else\unskip\hfill\fi\oprocendsymbol}

\usepackage{soul}

\usepackage{amsmath}
\usepackage{amsfonts}
\usepackage{amssymb}
\usepackage{amsthm}
\usepackage{mathtools}
\usepackage{bbm}
\usepackage{bm}
\usepackage{booktabs} 
\usepackage{graphicx}
\usepackage{float}
\usepackage[colorlinks,hidelinks]{hyperref}
\usepackage[dvipsnames]{xcolor}

\usepackage{cancel}
\usepackage[style=ieee, backend=biber]{biblatex}

\usepackage{geometry}
\usepackage{nicematrix}
\usepackage{algorithm}
\usepackage[noend]{algpseudocode}

\DeclareMathAlphabet{\mathbbold}{U}{bbold}{m}{n}
\newcommand{\one}{\mathbbold{1}}
\newcommand{\zero}{\mathbbold{0}}

\newcommand*\phantomrel[1]{\mathrel{\phantom{#1}}} 

\renewcommand{\j}{\mathbf{j}}

\newcommand{\inR}{\in \mathbb{R}}

\newcommand{\inZ}{\in \mathbb{Z}}

\newcommand{\thetas}{\theta_{\mathrm{s}}}
\newcommand{\goesto}{\rightarrow}

\newcommand{\transpose}[1]{{#1}^{\mathsf{T}}}

\newcommand{\pinv}[1]{{#1}^{\dagger}}

\newcommand{\image}[1]{\mathrm{im}\,#1}

\newcommand{\bcos}[1]{\mathbf{cos}#1}
\newcommand{\bsin}[1]{\mathbf{sin}#1}
\newcommand{\arcbsin}[1]{\mathbf{arcsin}#1}

\newtheorem{theorem}{Theorem}[]
\newtheorem{lemma}{Lemma}[]

\newtheorem{proposition}{Proposition}[]

\theoremstyle{definition}
\newtheorem{definition}{Definition}[]
\newtheorem{assumption}{Assumption}[]

\theoremstyle{remark}
\newtheorem{remark}{Remark}[]

\title{\bf A Fixed-Point Algorithm for the AC Power Flow Problem } 

\author{Liangjie Chen and John W. Simpson-Porco
    \thanks{The authors are with the Department of Electrical and Computer Engineering, University of Toronto, Toronto, ON, M5S 1A1 Canada. Email:
    \texttt{\footnotesize liangjie.chen@mail.utoronto.ca}, \texttt{\footnotesize jwsimpson@ece.utoronto.ca}.
    }
}

\begin{document}

\maketitle
\thispagestyle{empty}
\pagestyle{empty}

\begin{abstract}
This paper presents an algorithm that solves the AC power flow problem for balanced, three-phase transmission systems at steady state. 
The algorithm extends the ``fixed-point power flow'' algorithm in the literature
to include transmission losses, phase-shifting transformers, and a distributed slack bus model. The algorithm is derived by vectorizing the component-wise AC power flow equations and manipulating them into a novel equivalent fixed-point form. Preliminary theoretical results guaranteeing convergence are reported for the case of a two-bus power system.  We validate the algorithm through extensive simulations on test systems of various sizes under different loading levels, and compare its convergence behavior against those of classic power flow algorithms. 
\end{abstract}

\section{Introduction}
A fundamental problem which underpins many others in power system operations and control (e.g., optimal power flow, contingency analysis) is that of computing solutions to the power flow equations.
These equations describe the flow and balance of power in a synchronous AC power system at steady state \cite{overbye}, and are typically difficult to solve due to the inherently nonlinear relationship between power and voltage. Analytic solutions are rarely available, and solutions are instead computed via numerical methods. 

A plethora of heuristics and algorithms for analyzing power flow have been developed \cite{Molzahn1, Molzahn2, Bernie, convex-restriction}. The most standard approach to accurately solve the power flow problem is to use an iterative algorithm such as Newton-Raphson (NR) \cite[Chapter 6]{overbye}. 
The NR algorithm in particular is highly sensitive to initialization \cite{nr}, and heuristic or approximation procedures are sometimes needed to find suitable initial conditions \cite{Stott, CNN}. 
More generally, when iterative
algorithms fail, the failure mechanism can be difficult to determine, in that one cannot distinguish between poor initialization and an infeasible power flow case.
Thus, a robust algorithm for the power flow problem with transparent conditions for convergence is highly desirable. 

In recent years, there have been many fixed-point or contraction-based studies on the power flow problem and its solvability in various contexts (see, e.g., \cite{Cui, Bolognani, Wang, Bernstein}). These approaches provide sufficient conditions for the existence (and often, uniqueness) of a suitable power flow solution, and the contraction property naturally leads to a fixed-point iteration for computing that solution, which is guaranteed to converge from any initialization within the contraction region. Among these studies, our focus is power flow in balanced AC transmission systems. A recently developed research direction resulted in novel and particularly robust fixed-point based algorithms, termed fixed-point power flow (FPPF) \cite{FPPF1, FPPF2} and lossy DC power flow \cite{LossyDCPF}, to solve the power flow problem in the \textit{lossless} and  \textit{decoupled} contexts, respectively. While typical transmission lines have small losses, their presence can fundamentally change the physical behavior of the system and the solvability of the power flow problem \cite{Delabays, Stott}. Similarly, to accurately solve the full power flow problem, the coupling between active and reactive power cannot be ignored. Our goal here is to extend this line of research to incorporate network losses and other physically realistic modelling aspects not present in the original works, and derive an \textit{extended} FPPF algorithm to solve the AC power flow problem. 

\emph{Contributions:} The paper contains three main contributions. First, we develop a novel vectorization procedure for the AC power flow equations, which extends the vectorization developed in \cite{FPPF1,FPPF2} by incorporating resistive losses, phase-shifting transformers (PSTs), and a distributed slack bus (DSB) model \cite{distributed-slack}. Key to our vectorization is what we term the \textit{asymmetrically weighted (AW) incidence matrix} of the weighted bidirected graph describing the transmission grid. The proposed  algorithm is derived by exploiting a rank property of the AW incidence matrix. Second, we present preliminary theoretical results for a two-bus power system model, providing sufficient conditions for convergence of the algorithm towards the unique high-voltage power flow solution. Third and finally, we validate the algorithm via extensive numerical tests on standard power flow cases, and compare its performance to that of NR and the fast-decoupled (FDLF) method \cite{FDLF}. While our method is generally slower to converge than NR, it is more robust to changes in initialization in both lightly and heavily loaded networks. 
Due to space limitations, some proofs are omitted, but can be found in the thesis \cite{Thesis}. 

\textit{Notation}: We use $I_n$ to denote the $n\times n$ identity matrix, and use $\one_n, \zero_n, \zero_{n \times m}$ to denote the $n$-dimensional vector of all ones, zeros and the $n\times m$ zero matrix, respectively.\footnote{The subscripts are omitted when the the dimensions are easy to infer from the context.} 
We use $\transpose{M}, \pinv{M}$, and $M^{-1}$ to denote the transpose, left/right inverse, and inverse of a matrix $M$, respectively. 
Given an $x \inR^n$, $[x]$ denotes the diagonal matrix with $x$ on its main diagonal, and $x > \zero$ (resp. $x \geq \zero, x < \zero$) means that it is element-wise strictly positive (resp. nonnegative, strictly negative). In addition, $\bsin{(x)} \coloneqq \transpose{[\sin(x_1) \;\; \cdots \;\; \sin(x_n)]}$, with $\bcos{(x)}$, $\arcbsin{(x)}$ and $\sqrt{x}$ defined element-wise similarly. Given $u \inR^n$ and $v \inR^m$, $\mathrm{col}(u,v) \coloneqq \transpose{[u_1,\ldots,u_n,v_1,\ldots, v_m]} \inR^{n+m}$.

\section{Model Formulation} 
\subsection{Bidirected Graph Model of Transmission System}
\label{Sec:BiGraphModel}

To define the bidirected graph model used in this paper, we first review the standard notion of a weakly connected directed graph (digraph) and its properties \cite{bullo-networks}. 
A digraph is a pair $\mathcal{G} = (\mathcal{N}, \mathcal{E})$, with the node set $\mathcal{N} \coloneqq \{1, \ldots, n+m\}$, and the edge set $\mathcal{E} = \{e_1, \ldots, e_{\vert \mathcal{E}\vert}\} \subseteq \mathcal{N} \times \mathcal{N}$. The edge $e_k \coloneqq (i,j)$ models the connection between nodes $i$ and $j$, where one can travel from $i$ to $j$ (denoted by $i \goesto j$) but not vice versa. There is a weight function $W : \mathcal{E} \goesto [0,\infty)$ that equips each edge $e_k$ with a positive weight $w_k$ \cite{CLRS}.
A digraph is simple if there are no self loops, and it is weakly connected if there exists an \textit{undirected} path from any node to any other node in the graph. The \textit{incidence matrix} $A \inR^{(n+m) \times \vert \mathcal{E}\vert}$ of $\mathcal{G}$ is defined element-wise as $A_{ik} = 1$ if $(i,j) \in \mathcal{E}$, and $A_{ik} = -1$ if $(j,i) \in \mathcal{E}$; otherwise, $A_{ik} = 0$. Extracting only the $1$ and $-1$ entries, we can write $A = A^{+} - A^{-}$, where $A^+_{ik} = 1$ if and only if $A_{ik} = 1$, and $A^-_{ik} = 1$ if and only if $A_{ik} = -1$ \cite{FPPF1}, and allows us to define the \textit{undirected} incidence matrix $\vert A \vert$ as $\vert A \vert \coloneqq A^+ + A^-$. A \textit{cycle matrix} $C \inR^{\vert \mathcal{E}\vert \times n_c}$ has full column rank and satisfies $AC = \zero$, where $n_c = \vert \mathcal{E}\vert - (n+m-1)$ \cite{cycles}. 

In a transmission systems without PSTs, the physical branch models are symmetric with respect to changes in the current direction, resulting in a symmetric admittance matrix $Y$  \cite{MATPOWER}, and it suffices to model the circuit as a weighted digraph where each edge has one complex weight \cite{EleNetGraph}. However, PSTs create asymmetries in  $Y$, and we require a slightly more complex bidirectional digraph structure which assign different edge weights to each direction. A \emph{bidirected graph} is a digraph $\mathcal{G}_{\mathrm{b}} = (\mathcal{N}, \mathcal{E}_{\mathrm{b}})$ such that $(i,j) \in \mathcal{E}_{\mathrm{b}}$ if and only if $(j,i) \in \mathcal{E}_{\mathrm{b}}$ \cite{Delabays}. In transmission systems, a branch is a singular physical object (transmission line and/or transformer), so we use the symbol $e_k$ to denote both directions of a branch, i.e., $\{(i,j), (j,i)\}$ rather than just $(i,j)$ like in a digraph.\footnote{In order to not double-count, we still use $\vert \mathcal{E}\vert$ to denote the number of branches in the system.} Exactly one of the two elements  $\{(i,j), (j,i)\}$ is termed the ``forward'' edge and the other is the ``backward'' edge, and we may partition $\mathcal{E}_{\mathrm{b}}$ to be two disjoint sets $\mathcal{E}^+_{\mathrm{b}}, \mathcal{E}^-_{\mathrm{b}}$ representing the set of forward and backward edges.  Similar to the weight function of a digraph $\mathcal{G}$, there are the ``forward'' and ``backward'' weight functions $W^+, W^-$ mapping $\mathcal{E}^+_{\mathrm{b}}, \mathcal{E}^-_{\mathrm{b}}$ to $[0,\infty)$, respectively. For a branch $e_k$, we denote the $k$-th ``forward'' edge weight by $w_k^+ = W^+((i,j))$, and the $k$-th ``backward'' edge weight by $w_k^- = W^-((j,i))$; these weights can be collected into vectors $w^+, w^- \geq \zero$ respectively. With this definition, every bidirected graph $\mathcal{G}_{\mathrm{b}}$ induces a digraph $\mathcal{G}$, whose edge set is $\mathcal{E}^+_{\mathrm{b}}$ and weight function is $W^+$. We say $\mathcal{G}_{\mathrm{b}}$ is simple and weakly connected if $\mathcal{G}$ is simple and weakly connected. Throughout, we let $A$ and $C$ denote the incidence and cycle matrices of this induced forward-edge-only digraph $\mathcal{G}$ \cite{lawler}. 

A novel graph matrix that describes the structure of $\mathcal{G}_{\mathrm{b}}$ while accounting for the forward and backward edge weights $w^+, w^-$ is the \textit{asymmetrically-weighted (AW) incidence matrix} denoted by $\Gamma$, which is defined element-wise as
\begin{equation}
    \Gamma_{ik} \coloneqq 
    \left\{
    \begin{array}{cl}
        w_k^+ & \text{if } (i,j) \in \mathcal{E}^+_{\mathrm{b}}  \\
        -w_k^- & \text{if } (i,j) \in \mathcal{E}^-_{\mathrm{b}} \\
        0 & \text{otherwise}.
    \end{array}
    \right.
\end{equation}
One can quickly establish that $\Gamma = A^+[w^{+}] - A^-[w^{-}]$. Similar to the construction of undirected incidence matrix $\vert A \vert$, the \textit{undirected} AW incidence matrix is $\vert \Gamma \vert = A^+[w^{+}] + A^-[w^{-}]$. The following lemma forms a key step in the proof of Lemma \ref{lemma:M_B-full-row-rank} in Section \ref{ssec:fp}.
\begin{lemma}\label{lemma:same-sign}
    Suppose that $\mathcal{G}_{\mathrm{b}}$ is simple and weakly connected. Given $w^+, w^- > \zero$, if there exists a nonzero $x \inR^n$ such that $\,\transpose{\Gamma}x = \zero_{m}$, then either $x > \zero$ or $x < \zero$.
\end{lemma}
\begin{proof}
    See \cite[Lemma 2.1]{Thesis} for the proof.
\end{proof}

\subsection{Transmission System Power Flow Modelling}\label{subsec:transmission-model}
\begin{figure}[t]
    \centering
    \includegraphics[width=0.6\linewidth]{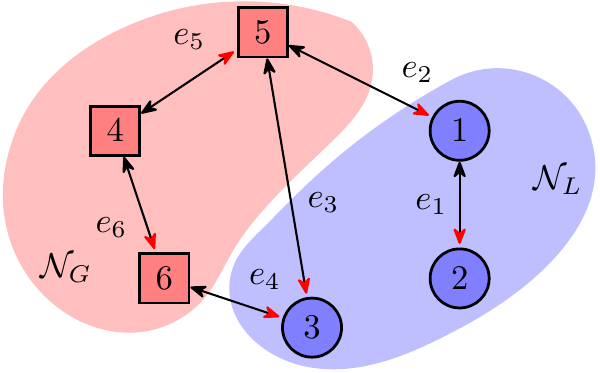}
    \caption{Bidirected graph model of a transmission system; forward and backward directions are marked in red and black, respectively.}
\vspace{-1.2em}
\label{fig:simple-example}
\end{figure}

We model a balanced transmission network at steady state as a simple, weakly connected bidirected graph discussed previously. The node set $\mathcal{N}$ models the buses, where $\mathcal{N}_L = \{1,...,n\}$ is the set of load (PQ) buses, $\mathcal{N}_G = \{n+1,...,n+m\}$ is the set of generator (PV) buses and $\mathcal{N} = \mathcal{N}_L \cup \mathcal{N}_G$. The edge set $\mathcal{E}_{\mathrm{b}}$ models the branches. Figure \ref{fig:simple-example} demonstrates an example. 
Each bus $i \in \mathcal{N}$ is associated with four physical quantities of interest: voltage magnitude $V_i$, phase angle $\theta_i$, real power injection $P_i$, and reactive power injection $Q_i$. 
For load buses, $P_i, Q_i$ are known and $V_i, \theta_i$ are unknown. For generator buses, $P_i, V_i$ are known and $Q_i, \theta_i$ are unknown. 
We can vectorize the quantities associated with each bus and partition the vectors according to the $\mathcal{N}_L, \mathcal{N}_G$ subsets as $P = \mathrm{col}(P_L, P_G)$, $Q = \mathrm{col}(Q_L, Q_G)$, and $V = \mathrm{col}(V_L, V_G)$,
where $P, Q_L, V_G$ are known, and $Q_G, V_L$ and $\theta$ are unknown. In particular, the goal of the power flow study is to solve for voltage magnitudes $V_L$ and phases $\theta$.\footnote{In this study we do not consider generator reactive power limits; see \cite[Example 6.11]{overbye} for further information on this case.}

Each branch is modelled with a PST (located at the side of bus $i$ if $(i,j) \in \mathcal{E}^+_{\mathrm{b}}$) connected to a $\Pi$-model transmission line \cite[Figure 3.1]{MATPOWER}.
The transmission line has a series admittance $y = g - \j b$, where $g \leq 0$ is the conductance and $b > 0$ is the susceptance, and a shunt susceptance $b_c$. The transformer has a complex turns ratio $\tau = t \exp(\j\thetas)$ where $t, \thetas$ respectively represent the tap ratio and phase shift. The conductance and susceptance matrices are denoted by $G$ and $B$, respectively, such that $Y = G  + \j B$. When there is at least one PST in the system, i.e., a transformer with nonzero $\thetas$, both $G$ and $B$ become asymmetric. We partition the $B$ matrix based on the $\mathcal{N}_L, \mathcal{N}_G$ subsets as 
\[
    B = 
    \left[    
    \begin{array}{c|c}
        B_{LL} & B_{LG} \\
        \hline
        B_{GL} & B_{GG}
    \end{array}
    \right].
\]
\vspace{-1.5em}
\begin{assumption}\label{ass:B}
The sub-matrix $B_{LL}$ is strictly diagonally dominant. If bus $i$ is connected to bus $j$, then $B_{ij} > 0$.
\end{assumption}
Strict diagonal dominance implies that $B_{LL}$ is invertible \cite[Theorem 6.1.10]{Horn1}, and physically means that capacitive shunt elements  do not overcompensate the inductive network \cite[Assumption 2.2]{FPPF1}. The off-diagonal elements that correspond to network branches being positive requires that as the angle $\thetas$ of the PST increases, the corresponding branch $R/X$ ratio must be sufficiently small such that $b/g > \tan \thetas$ \cite[Assumption 2.2]{Thesis}. Together, Assumption \ref{ass:B} also implies that $-B_{LL}^{-1}$ is a nonnegative matrix \cite[Theorem 2.3 $\mathrm{N}_{38}$]{Berman}.

\section{The Extended Fixed-Point Power Flow Algorithm}

\subsection{A Novel Vectorization of the Power Flow Equations}

Our development begins from the standard power flow equations \cite{overbye} with the DSB model \cite{distributed-slack}:
\begin{subequations}\label{eq:PQ-general}
\begin{align}
    P_i &= V_i\sum_{j=1}^{n+m} V_j(G_{ij}\cos\varphi_{ij} + B_{ij}\sin\varphi_{ij}), \quad i \in \mathcal{N} \label{eq:P-general} \\
    Q_i &= V_i\sum_{j=1}^{n+m} V_j(G_{ij}\sin\varphi_{ij} - B_{ij}\cos\varphi_{ij}), \quad i \in \mathcal{N}_L  \label{eq:Q-general} 
\end{align}
\end{subequations}
where $\varphi_{ij}$ is the branch-wise phase difference $\theta_i - \theta_j$. In a DSB model, the nodal real power injection $P_i = \bar{P_i} + \alpha_iP_{\mathrm{s}}$ is the sum of the \textit{known} injection $\bar{P_i}$ and a portion of the \textit{unknown} slack power $P_{\mathrm{s}}$. The known constants $\alpha_i$ are called \textit{participation factors}, such that $\alpha_i > 0$ for all the generator buses in the distributed slack bus subset of $\mathcal{N}_G$ and $0$ otherwise, subject to $\sum_i \alpha_i = 1$.  

The goal of this subsection is to vectorize \eqref{eq:PQ-general}, which will involve a change of variables. We first define the \textit{open-circuit load voltage} $V_{L}^{\circ} \inR^{n}$ as $V_{L}^{\circ} = -B_{LL}^{-1}B_{LG}V_G$, which is fixed since $B_{LL}, B_{LG}$ and $V_G$ are known, and one can show that $V_L^{\circ} > \zero$ \cite{FPPF1}. Using this quantity and the generator voltage magnitudes $V_G$, we let $V^{\circ} \coloneqq \mathrm{col}(V_L^{\circ},V_G)$,
and define the \textit{normalized} load voltage as
\begin{equation}\label{eq:v-def}
    v \coloneqq [V_{L}^{\circ}]^{-1}V_{L} \inR^n,
\end{equation} 
and extend this normalization scheme for all $V_i$'s as
\begin{equation}\label{eq:gv}
    g(v) \coloneqq [V^{\circ}]^{-1}V = \begin{bmatrix}
        v \\ \one_{m}
    \end{bmatrix} \iff V = [V^{\circ}]g(v).
\end{equation} 

\begin{definition}[Branch stiffness matrices]\label{def:DBDG}
Given $V^{\circ}$ and the matrices $G$ and $B$, the \emph{branch stiffness matrices} are the following diagonal matrices
\begin{equation*}\label{eq:stiffness-mats}
\left.
\begin{aligned}
    D_{G}^{+} &= \left[V_i^{\circ}V_j^{\circ}G_{ij}\right]_{(i,j) \in \mathcal{E}^+_{\mathrm{b}}} \;\; 
    D_{G}^{-} = \left[V_i^{\circ}V_j^{\circ}G_{ji}\right]_{(j,i) \in \mathcal{E}^-_{\mathrm{b}}} 
    \\
    D_{B}^{+} &= \left[V_i^{\circ}V_j^{\circ}B_{ij}\right]_{(i,j) \in \mathcal{E}^+_{\mathrm{b}}} \;\; 
    D_{B}^{-} = \left[V_i^{\circ}V_j^{\circ}B_{ji}\right]_{(i,j) \in \mathcal{E}^-_{\mathrm{b}}}.
\end{aligned}
\right.
\end{equation*}
\end{definition}
These matrices are extensions of those defined in \cite{FPPF1};  the diagonal elements form the forward and backward edge weights in our bidirected graph model of the transmission system. Consequently, following Section \ref{Sec:BiGraphModel}, we define the following asymmetrically weighted incidence matrices as 
\begin{equation}\label{eq:FourGamma}
\left.
\begin{aligned}
\Gamma_B \coloneqq A^{+}D_{B}^{+} - A^{-}D_{B}^{-}  \quad \vert \Gamma_B \vert \coloneqq  A^{+}D_{B}^{+} + A^{-}D_{B}^{-} \\
\Gamma_G \coloneqq A^{+}D_{G}^{+} - A^{-}D_{G}^{-} \quad \vert \Gamma_G \vert \coloneqq  A^{+}D_{G}^{+} + A^{-}D_{G}^{-}.
\end{aligned}
\right.
\end{equation}
Lastly, we define the following nonlinear map \cite{FPPF1}
\begin{equation}\label{eq:h-full}
    h(v) = 
    \left[\transpose{(A^+)} g(v)\right]\transpose{(A^-)} g(v),
\end{equation}
which transforms the nodal normalized voltage magnitudes $g(v)$ to branch-wise normalized voltage magnitude \emph{products} $h(v)$. One can easily verify that the diagonal elements of $D_B^{+}[h(v)], D_B^{-}[h(v)]$ are the $V_iV_jB_{ij}, V_iV_jB_{ji}$ terms in \eqref{eq:PQ-general}, respectively, and similarly for $D_G^{+}[h(v)], D_G^{-}[h(v)]$. These elements are associated with the \textit{branches}, and multiplication by matrices in \eqref{eq:FourGamma} maps them into the \textit{nodal} quantities captured by the summation in \eqref{eq:PQ-general}. 

We are now ready to vectorize the power flow equations \eqref{eq:PQ-general}. We begin with the active power flow equation \eqref{eq:P-general}, and expand its right-hand side to be
\begin{equation}\label{eq:P-general-long}
    V_i^{2}G_{ii} + 
    \sum_{j\neq i}  V_iV_jG_{ij}\cos\varphi_{ij} +
    \sum_{j\neq i}  V_iV_jB_{ij}\sin\varphi_{ij}.
\end{equation}
For the first term in \eqref{eq:P-general-long}, we extract the diagonal entries of $G$ into the diagonal matrix $[G_{ii}]$. Applying \eqref{eq:gv}, we vectorize the $V_i^2G_{ii}$ terms as $[V^{\circ}][g(v)][G_{ii}][V^{\circ}]g(v)$. Using the incidence matrix $A$, we vectorize the phase differences $\varphi_{ij}$ in \eqref{eq:P-general-long} as $\transpose{A}\theta$, and  
we define the following change of variable
\begin{equation}\label{eq:psi-def}
    \psi \coloneqq \bsin{(\transpose{A}\theta)} \implies \bcos(A^{\sf T}\theta) = \sqrt{\one_{\vert \mathcal{E}\vert} - [\psi]\psi}.
\end{equation}
Applying the asymmetrically weighted incidence matrices in \eqref{eq:FourGamma}, we can verify that 
\begin{equation}\label{eq:P-vector-noR}
\begin{aligned}
    P &= [V^{\circ}][g(v)] [G_{ii}][V^{\circ}] g(v) 
    + 
    \vert \Gamma_G \vert[h(v)]\sqrt{\one_{\vert \mathcal{E}\vert} - [\psi]\psi}  
    \nonumber \\ 
    &\phantomrel{=}{} 
    + \Gamma_B[h(v)]\psi
    \end{aligned}
\end{equation}
is the vectorization of \eqref{eq:P-general} by expanding right-hand side. As mentioned at the beginning of this section, we can write the nodal power injection as $P = \bar{P} + P_{\mathrm{s}}\alpha$. 
In power flow computations with a single slack bus, we solve for $n+m-1$ phase angles and $n$ load voltage magnitudes using $2n+m-1$ power flow equations since the slack bus with fixed voltage is chosen to compensate for the unknown $P_{\mathrm{s}}$ and its corresponding active power flow equation is removed. With the DSB model, a slightly more nuanced procedure is required to eliminate this degree of freedom. 
To this end, 
since
$\sum \alpha_i = 1$, we can construct a full column rank matrix $R \inR^{(n+m) \times (n+m-1)}$ such that $\transpose{R}\alpha = \zero$. 
Left-multiplying \eqref{eq:P-vector-noR} by $R^{\sf T}$, we eliminate $P_{\rm s}$ and obtain
\begin{align}
    \transpose{R}\bar{P}
    &= \transpose{R}\big(
        [V^{\circ}][g(v)] [G_{ii}][V^{\circ}] g(v) 
        \nonumber \\
        &+
        \vert \Gamma_G \vert[h(v)]\sqrt{\one_{\vert \mathcal{E}\vert} - [\psi]\psi}  
        + 
        \Gamma_B[h(v)]\psi 
    \big).  \label{eq:P-vector}
\end{align}
which is now a system of $n+m-1$ active power flow equations. This reduction is a generalization of the standard single-slack-bus elimination procedure for power flow studies \cite[Chapter 3.1]{Thesis}. 
Once $\psi$ and $v$ are known, $P_{\mathrm{s}}$ can be uniquely recovered. Denoting the right-hand side of \eqref{eq:P-general} by $\mathcal{P}(\psi, v)$, since $\sum_i \alpha_i = \transpose{\one}\alpha = 1$, from \eqref{eq:P-general} we have
\[ 
P_{\mathrm{s}} \alpha = \bar{P} - \mathcal{P}(\psi,v)\,\, \implies\,\,
P_{\mathrm{s}} = \transpose{\one}(\bar{P} - \mathcal{P}(\psi,v)).
\]

We now proceed to vectorize the $n$ reactive power flow equations \eqref{eq:Q-general} corresponding to $Q_L \inR^n$, the known load reactive power injections. We extract the $n \times n$ submatrix of $[B_{ii}]$, and the top $n \times \vert \mathcal{E} \vert$ submatrices of $\Gamma_G, \vert \Gamma_B \vert$, and denote them with the additional $\cdot_L$ subscript. We may then write the vectorized reactive power flow equation as
\begin{align}
    Q_L = 
    &-[V_{L}^{\circ}][v][B_{ii}]_{L}[V_{L}^{\circ}]v + 
    \Gamma_{G_L}[h(v)]\psi \nonumber \\ 
    &- 
    \vert \Gamma_{B_L} \vert [h(v)]\sqrt{\one_{\vert \mathcal{E}\vert} - [\psi]\psi}. \label{eq:Q-vector-1}
\end{align}
Finally, we define the invertible $n \times n$ \textit{nodal stiffness matrix} 
\[
    S \coloneqq \tfrac{1}{4}[V_{L}^{\circ}]B_{LL}[V_{L}^{\circ}],
\]
and by \cite[Lemma A.3]{FPPF1}, the vectorized reactive power flow equation \eqref{eq:Q-vector-1} is equivalent to 
\begin{align}
    Q_L &= 4[v]S(\one_{n} - v)
    + 
    \Gamma_{G_L}[h(v)]\psi  \nonumber \\
    &\phantomrel{=}{} + 
    \vert \Gamma_{B_L} \vert [h(v)]\left(\one_{\vert \mathcal{E}\vert} - \sqrt{\one_{\vert \mathcal{E}\vert} - [\psi]\psi} \right).\label{eq:Q-vector-2}
\end{align} 

\subsection{Fixed-Point Reformulation of Power Flow Equations}\label{ssec:fp}

To derive the proposed algorithm, we will manipulate the vectorized power flow equations \eqref{eq:P-vector}, \eqref{eq:Q-vector-2} into a fixed-point form. We begin with \eqref{eq:P-vector}, and define
\[
    M_B = \transpose{R}\Gamma_B \inR^{(n+m-1) \times \vert \mathcal{E} \vert},
\]
which can be interpreted as a ``reduced'' version of the AW incidence matrix $\Gamma_B$. The following lemma is a key result used to show the equivalence between the fixed-point reformulation and the standard power flow equations \eqref{eq:PQ-general}.

\begin{lemma}\label{lemma:M_B-full-row-rank}
    The matrix $M_B$ has full row rank. 
\end{lemma}

\begin{proof} 
    We equivalently prove that $\transpose{M}_B$ has a trivial kernel. By contradiction, assume that there exists a nonzero $x \in \ker \transpose{M}_B$. Then either
    \begin{enumerate}
        \item \label{item:MB-full-rank-1} $x \in \ker R$, or
        \item \label{item:MB-full-rank-2} there exists some $y = Rx$ such that $y \in \ker\transpose{\Gamma}_B$.
    \end{enumerate}
    By construction, $R$ has full column rank, so case \ref{item:MB-full-rank-1}) cannot occur. For case \ref{item:MB-full-rank-2}), suppose that such $y$ exists and is nonzero. Denote the $k$-th column of $\Gamma_B$ by $r_k$, then $\transpose{\Gamma}_B y = \zero$ if and only if $\langle r_k, y \rangle= 0$ for all $k = 1, ..., \vert \mathcal{E} \vert$. By the construction of the branch stiffness matrices $D_B^{+},D_B^{-}$ and Assumption \ref{ass:B}, for each $k \in \{1,\ldots,\vert \mathcal{E}\vert\}$, $r_k$ contains exactly two nonzero elements at $w_{k,i} = V_i^{\circ}V_j^{\circ}B_{ij}$ and $w_{k,j} = -V_i^{\circ}V_j^{\circ}B_{ji}$,
    so $\transpose{\Gamma}_B y = \zero$ if and only if $\langle r_k, y \rangle = V_i^{\circ}V_j^{\circ}(B_{ij}y_i - B_{ji}y_j) = 0$ for all $k = 1, ..., \vert \mathcal{E}\vert$. However, since $V_i^{\circ}, V_j^{\circ} > 0$ and $B_{ij}, B_{ji} > 0$, $y_i, y_j$ must be both positive, both negative, or both zero. Lemma \ref{lemma:same-sign} implies that $y > \zero$ or $y < \zero$, so $\alpha_i \geq 0$ implies that $\transpose{y}\alpha \neq 0$ always holds. Finally, by the construction of $R$, $\transpose{y}\alpha \neq 0$ implies that $y \notin \image R$, i.e., there does not exist a nonzero $x$ such that $y = Rx \in \ker \transpose{\Gamma}_B$ and case \ref{item:MB-full-rank-2}) cannot hold, which completes the proof.
\end{proof}

Let $K$ be a matrix whose columns form a basis of $\ker M_B$, so $M_BK = \zero$. By the rank-nullity theorem and Lemma \ref{lemma:M_B-full-row-rank}, $K$ must have $n_c$ linearly independent columns.
In addition, Lemma  \ref{lemma:M_B-full-row-rank} implies that matrix $M_B$ has a right inverse denoted by $\pinv{M}_B$, where $M_B\pinv{M}_B = I_{n+m-1}$. Finally, notice that \eqref{eq:P-vector} is linear in $\psi\inR^{\vert \mathcal{E}\vert }$ in the last term if we know $v$ and the square root term. Since $v > \zero$, $[h(v)]^{-1}$ exists, so we can rearrange \eqref{eq:P-vector} to obtain
\begin{align}
    \psi &= f_P(\psi,v,x_c) 
    \nonumber \\
    &\!\coloneqq [h(v)]^{-1}\pinv{M}_B \transpose{R}\left(\bar{P} - 
        [V^{\circ}][g(v)] [G_{ii}][V^{\circ}] g(v) \right. \nonumber \\ 
    &\phantomrel{=}{} - \left.
        \vert \Gamma_G \vert [h(v)]
        \sqrt{\one_{\vert \mathcal{E}\vert} - [\psi]\psi}  
    \right) + [h(v)]^{-1}Kx_c, \label{eq:psi-solution-full}
\end{align} 
where the final term $[h(v)]^{-1}Kx_c$ characterizes the homogeneous part of the solution for $\psi$ with an additional variable $x_c \inR^{n_c}$. While \eqref{eq:psi-solution-full} results in the voltage phase solution in terms of $\psi$, we ultimately want the bus voltage phase $\theta \inR^{n+m}$. To recover $\theta$ from $\psi$, note that for any $k \inZ$, $\bsin{(\transpose{A}\theta + 2\pi k)} = \psi$ and $\transpose{C}k$ has integer elements \cite{FPPF1}. Recalling the property that $AC = \zero$, we must have $\transpose{C}\arcbsin{(\psi)} = \transpose{C} \left(\transpose{A}\theta  + 2\pi k \right) = 2\pi \transpose{C}k$, which results in the ``loop-flow'' constraint
\begin{equation}\label{eq:xc-solution-full}
    \transpose{C}\arcbsin{(\psi)} \bmod 2\pi =  \zero_{n_c}.
\end{equation}

Next, we manipulate the reactive power flow equation \eqref{eq:Q-vector-2} into a fixed-point form by 
left-multiplying both sides of the equation by $\frac{1}{4}S^{-1}[v]^{-1}$ to obtain
\begin{align}
    v = f_Q(\psi,v)
    &\coloneqq
    \one_n - \frac{1}{4}
    S^{-1}[v]^{-1}
    \Big(\left(Q_L - 
    \Gamma_{G_L}[h(v)]\psi \right) \nonumber \\
    &\!\!\!\!\!\!-
    \vert \Gamma_{B_L} \vert [h(v)]\left(\one_{\vert \mathcal{E}\vert} - \sqrt{\one_{\vert \mathcal{E}\vert} - [\psi]\psi} \right)
    \Big). 
    \label{eq:v-solution-full}
\end{align}
We summarize our above development in the following theorem, which is the main theoretical result of this paper. 
\begin{theorem}[\bf FPPF]\label{Thm:Main}
    Consider the normalized load voltage magnitudes $v \inR^{n}$ defined in \eqref{eq:v-def}, a vector $x_c \inR^{n_c}$ and the change of variable $\psi \coloneqq \bsin( \transpose{A}\theta)$ defined in \eqref{eq:psi-def}. The following statements are equivalent: 
    \begin{enumerate}
        \item[(i)] $(\theta, V_L)$ solves the vectorized power flow equations \eqref{eq:P-vector}, \eqref{eq:Q-vector-2};
        \item[(ii)] $(\psi, v, x_c)$ satisfy the fixed point equations \eqref{eq:psi-solution-full},  \eqref{eq:v-solution-full}, and the loop flow constraint \eqref{eq:xc-solution-full}. 
    \end{enumerate}
\end{theorem}

When there are no network losses and PSTs in the system, Theorem \ref{Thm:Main} recovers \cite[Theorem 3.5]{FPPF1}. If the system is radial, i.e., $\vert \mathcal{E}\vert = n+m-1$, then the loop flow constraint \eqref{eq:xc-solution-full} and the homogeneous solution in \eqref{eq:psi-solution-full} are not required since the kernel of the incidence matrix $A$ is trivial \cite{bullo-networks}, so $n_c = 0$ and we no longer need the variable $x_c$.

\subsection{The Extended FPPF Algorithm}

Based on Theorem \ref{Thm:Main}, we propose the following extended fixed-point power flow algorithm, which will be tested extensively in Section \ref{sec:num}.

\begin{algorithm}[H]
\caption{The Fixed-Point Power Flow Algorithm}\label{alg:1}
\begin{algorithmic}
\Require Power flow data, tolerance $\epsilon$, max. iterations $L$

\State $v^k \leftarrow V_L/V^{\circ}_L$
,  $\psi^k \leftarrow \bsin{(\transpose{A}\theta)}$
,  $x_c^k \leftarrow \zero_{n_c}$
\State $k \gets 0$

\State Compute power balance mismatch with $\psi^k, v^k$ 
\While{$\text{mismatch} > \epsilon \text{ and } k < L$}
\State $v^{k+1} \leftarrow f_Q(\psi^k, v^k, x_c^k$)
\If{$n_c > 0$}
    \State $x_c^{k+1} \leftarrow \text{ Newton step on } \transpose{C}\arcbsin{(\psi^k)} =  \zero_{n_c}$
\EndIf
    \State $\psi^{k+1} \leftarrow f_P(\psi^k, v^{k+1}, x_c^{k+1})$
    \State Compute power balance mismatch with $\psi^{k+1}, v^{k+1}$
    \State $k \gets k + 1$
\EndWhile

\noindent \Return $\psi^{k+1}$, $v^{k+1}$, power balance mismatch
\end{algorithmic}
\end{algorithm}

The power flow data consist of the matrices and vectors related to the network topology, admittance matrix, loading profiles and other constants used in the reformulation; see \cite[Section 6.1]{Thesis} for the discussion on their construction. The Newton step to evaluate $x_c^{k+1}$ is
    \[
        x_c^{k+1}  = x_c^k - \left(J_c^{k}\right)^{-1}\transpose{C}\arcbsin{(\psi^{k})},
    \]
    where
    $
        J_c^{k} \coloneqq \transpose{C}\left(I_{\vert \mathcal{E} \vert} - [\psi^k]^2\right)^{-1/2} [h(v^{k+1})]^{-1} K
    $
    is the Jacobian matrix of the constraint evaluated at $\psi^k$; note that $J_c^k$ is computed using the most up-to-date $v^{k+1}$. 
    
\begin{remark}[\bf Update order] \label{rmk:update-order}
    In the spirit of \cite{FFW}, we use the most updated version of a variable to evaluate the update of the other variables. For the numerical simulations in Section \ref{sec:num}, we follow a ``$v$-$x_c$-$\psi$'' order: we first compute $v^{k+1}$ with $\psi^k, v^k, x_c^k$, then compute $x_c^{k+1}$ with $v^{k+1}$ instead of $v^k$, and finally $\psi^{k+1}$ with both $v^{k+1}$ and $x_c^{k+1}$. Other orders are also possible; see \cite[Section 6.2.1]{Thesis} for a detailed discussion on the effect of different update orders. \hfill \oprocend
\end{remark}

\section{Analysis of Algorithm for Two-Bus System}
As preliminary theoretical analysis of our approach, we conduct a convergence analysis of the FPPF algorithm on the two-bus power flow problem \cite{vancutsem}. Since a fixed-point algorithm like ours naturally leads to a contraction analysis, we re-frame the problem as one of constructing a compact invariant set on which the FPPF algorithm is a contraction. 

\subsection{Problem Setup}
\begin{figure}[htb]
    \centering
    \includegraphics[width=0.6\linewidth, trim=3pt 4pt 3pt 4pt,clip]{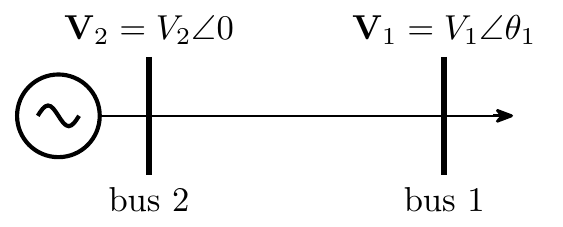}
    \caption{Two bus system}
    \label{fig:2bus}
\end{figure}

Consider the two-bus model in Figure \ref{fig:2bus}, where bus 2 is the only PV/slack bus, bus 1 is the PQ bus, and the branch parameters $t, \thetas, g, b, b_c$ are as described in Section \ref{subsec:transmission-model}. 

Let $\bar{t} \coloneqq t -1$, we define the following system constants
\[
    \Tilde{g} \coloneqq \dfrac{g\cos\thetas - b\sin\thetas}{\bar{t}+1}, \,
    \Tilde{b} \coloneqq \dfrac{b\cos\thetas + g\sin\thetas}{\bar{t}+1}, \,
    \hat{b} \coloneqq b - \dfrac{b_c}{2}. 
\]
Next, let $\rho \coloneqq {g}/{\hat{b}}, \tilde{\rho} \coloneqq  {\tilde{g}}/ {\tilde{b}}$ be two constants that quantify the system $R/X$ ratio; realistically, $\rho \geq 0$, and Assumption \ref{ass:B} implies $\tilde{\rho} \geq 0$. We define real and reactive power loading margins as 
\[
    \Tilde{\gamma}_P \coloneqq \dfrac{\bar{P}_1 }{\tilde{b} V_1^{\circ}V_2^{\circ}}, \quad \Tilde{\gamma}_Q \coloneqq  \dfrac{Q_1}{\tilde{b} V_1^{\circ}V_2^{\circ}},
\]
where the open-circuit load voltage evaluates to be $V_1^{\circ} = ({\Tilde{b}}/{\hat{b})}V_2$. Making the change of variable $x \coloneqq v -1$ and defining the state vector $\xi \coloneqq (\psi, x)$, we can write the FPPF algorithm update rule as
\begin{equation} \label{eq:2bus-fmu}
    \xi_{k+1} =
    \underbrace{{
    \renewcommand{\arraystretch}{2}
    \begin{bmatrix}
        \dfrac{-\tilde{\gamma}_P}{x_k+1} + \rho \left(x_k+1\right) - \tilde{\rho}\sqrt{1 - \psi_k^2} \\
        \dfrac{\tilde{\gamma}_Q}{x_k+1} - \tilde{\rho} \psi_{k+1} + \sqrt{1 - \psi_{k+1}^2} - 1
    \end{bmatrix}
    }.}_{F_{\mu}(\xi_k)}
\end{equation} 
The $\mu$ subscript in \eqref{eq:2bus-fmu} refers to the vector of ``perturbations''
\[
    \mu \coloneqq \transpose{[g \quad b_c \quad \bar{t} \quad \thetas ]}.
\]
If $\mu = \zero$, then the system is lossless, the transformer is absent, and the branch is simply a series reactance given by $1/b$; in this case, $\rho = \tilde{\rho} = 0$. We call this case the ``nominal system'' and denote the corresponding mapping that defines the update rule by $F_{\zero}$, otherwise we call it the ``full system''. 

The ultimate goal of the ensuing analysis is to (i) construct a $F_{\mu}$-invariant set for \eqref{eq:2bus-fmu}, and (ii) provide sufficient conditions that $F_{\mu}$ is a contraction on this set; this will guarantee convergence of \eqref{eq:2bus-fmu} to a (locally) unique solution. Our approach will be to develop results for $F_\zero$, and then extend these results to $F_\mu$ for sufficiently small $\mu$ values.

\subsection{Nominal System Results}  

When $\mu = \zero$, $\hat{b} = \tilde{b} = b > 0$, $\tilde{g} = g = 0$, and $V_1^{\circ} = V_2^{\circ}$. We denote the reduced constants $\tilde{\gamma}_P, \tilde{\gamma}_Q$ by $\gamma_P, \gamma_Q$, respectively. The assumption below provides a crucial characterization of permissible loading margins.  

\begin{assumption}\label{ass:2bus-f0-pq-bound}
$0 < 4\gamma_P^2 - 4\gamma_Q  < 1$.
\end{assumption}

Assumption \ref{ass:2bus-f0-pq-bound} is a standing assumption for the rest of this section. It states that the load is inductive, and restricts the amount of loading in the system\footnote{It is physically valid for $\gamma_P,\gamma_Q$ to be both zero, but we are interested in the case of a strictly positive active power injection at the generator bus.}. Now, define the compact and convex set 
\begin{equation}\label{eq:2bus-f0-Aset}
\mathcal{A}(k_1, k_2) \coloneqq \{\xi :
\vert \psi \vert \leq k_1, \vert x \vert \leq k_2\},
\end{equation}
which is a closed box in $\mathbb{R}^2$ centered at the origin, parameterized by some $k_1, k_2 > 0$ to be determined. 

\begin{theorem}\label{thm:2bus-f0-inv}
Let $k_1^{-} \coloneqq -{\gamma_P}/({ 1-k_2^{-} })$, where
\[
\begin{aligned}
k_2^- &\coloneqq 1 - \sqrt{\frac{1}{2}  + \gamma_Q + \sqrt{\frac{1}{4} + \gamma_Q - \gamma_P^2}}, \nonumber \\
k_2^+ &\coloneqq 1 - \sqrt{\frac{1}{2}  + \gamma_Q - \sqrt{\frac{1}{4} + \gamma_Q - \gamma_P^2}}.
\end{aligned}
\]
The set $\mathcal{A}(k_1, k_2)$ is $F_{\zero}$-invariant for any  $k_1 \in [k_1^{-},1]$ and $k_2 \in [k_2^{-} , k_2^{+}]$.
\end{theorem}
\begin{proof}
    See \cite[Section 5.2.1]{Thesis} for the proof.
\end{proof}
We are especially interested in the \textit{smallest} $F_{\zero}$-invariant set denoted by $\mathcal{A}^{-} \coloneqq \mathcal{A}(k_1^-, k_2^-)$. In fact, one can verify by direct substitution that $\xi = \transpose{[k_1^{-} \; -k_2^{-}]}$ is the desired high-voltage solution, and it exists on the boundary of $\mathcal{A}^-$. 

\begin{theorem}\label{thm:2bus-f0-contraction}
    $F_{\zero}$ is a contraction on $\mathcal{A}^-$ in the $\ell_{\infty}$ norm.
\end{theorem}
\begin{proof}
    See \cite[Section 5.2.2]{Thesis} for the proof.
\end{proof}

The above results implicitly recover the calculations in \cite[Chapter 2]{vancutsem}, along with the existence/uniqueness result of \cite{FPPF2} when restricted to the two-bus case.

\subsection{Full System Analysis}
We now return to the update rule \eqref{eq:2bus-fmu} for $\mu \neq \zero$. First, note that we can write $\tilde{\gamma}_P = k_{\mu}\gamma_P$ and $\tilde{\gamma}_Q = k_{\mu}\gamma_Q$, where $k_{\mu} \coloneqq ({b\hat{b}}) / {\tilde{b}^2}$. Let $\epsilon_1 = \epsilon_1(\mu, \gamma_P, \gamma_Q)$ and $\epsilon_2 = \epsilon_2(\mu, \gamma_P, \gamma_Q)$ be nonnegative constants to be determined, and define the set 
\[
\mathcal{A}_{\epsilon}(k_1, k_2) \coloneqq \mathcal{A}(k_1+\epsilon_1, k_2+\epsilon_2),
\]
then we can define $\mathcal{A}^{-}_{\epsilon} \coloneqq \mathcal{A}_{\epsilon}(k_1^{-}+\epsilon_1, k_2^{-}+\epsilon_2)$. 
Intuitively, this set is ``slightly expanded'' from $\mathcal{A}^{-}$  when  $\mu \neq \zero$.  
We now derive a condition on $\epsilon \coloneqq (\epsilon_1, \epsilon_2)$ such that $\mathcal{A}^{-}_{\epsilon}$ is $F_{\mu}$-invariant, which is true if and only if for every $\xi_k \in \mathcal{A}^{-}_{\epsilon}$,
\[
\begin{aligned}
    \vert \psi_{k+1} \vert
    &= \left \vert 
        \dfrac{-k_{\mu}\gamma_P}{x_k+1} + \rho \left(x_k+1\right) - \tilde{\rho}\sqrt{1 - \psi_k^2}
    \right \vert
    \leq k_1^{-} + \epsilon_1, \\
    \vert x_{k+1} \vert 
    &= \left \vert
        \dfrac{k_{\mu}\gamma_Q}{x_k+1} - \tilde{\rho} \psi_{k+1} + \sqrt{1 - \psi_{k+1}^2} - 1    
    \right \vert 
    \leq k_2^{-} + \epsilon_2.
\end{aligned}
\]
By triangle inequality and the fact that $\vert x_k \vert \leq k_2^{-}$, the $\vert \psi_{k+1} \vert$ term in first inequality can be upper bounded by 
\[ 
    \dfrac{-k_{\mu}\gamma_P}{1-k_2^{-}-\epsilon_2}  +  \rho\left(1+k_2^{-}+\epsilon_2\right) + \tilde{\rho}.
\]
Adding and subtracting a $k_1^- = \gamma_P / (1-k_2^-)$ in the expression above, if $\epsilon$ satisfies
\begin{subequations}\label{eq:2bus-fmu-invariance-eps}
\begin{equation}\label{eq:2bus-fmu-invariance-eps1}
    \dfrac{-k_{\mu}\gamma_P}{1-k_2^{-}-\epsilon_2} + \frac{\gamma_P}{1-k_2^{-}} + \rho\left(1+k_2^{-}+\epsilon_2\right) + \tilde{\rho}  \leq \epsilon_1,
\end{equation}
then any $\vert \psi_{k}\vert \leq k_1^{-}+\epsilon_1$ implies $\vert \psi_{k+1}\vert \leq k_1^{-}+\epsilon_1$.
Note that
we also need $\epsilon_2 < 1-k_2^{-}$ to prevent division by zero in \eqref{eq:2bus-fmu-invariance-eps1}. 
Similarly, the $\vert x_{k+1} \vert$
term in the second inequality can be upper bounded by 
\[
\begin{aligned}
\dfrac{-k_{\mu}\gamma_Q}{1-k_2^{-}-\epsilon_2}
    + \tilde{\rho} \left(k_1^{-}+\epsilon_1\right)
    - \sqrt{1 - \left(k_1^{-}+\epsilon_1\right)^2}+ 1.
\end{aligned}
\]
Adding and subtracting a $k_2^-$ above, if $\epsilon$ satisfies
\begin{align}
    \dfrac{-k_{\mu}\gamma_Q}{1-k_2^{-}-\epsilon_2}
    &+ \left(1 - k_2^{-} \right)
    + \tilde{\rho} \left(k_1^{-}+\epsilon_1\right) 
    \nonumber \\
    &\phantomrel{=}{}
    - \sqrt{1 - \left(k_1^{-}+\epsilon_1\right)^2}
    \leq \epsilon_2, \label{eq:2bus-fmu-invariance-eps2}
\end{align}
\end{subequations}
then any $\vert x_{k}\vert \leq k_2^{-}+\epsilon_2$ implies $\vert x_{k+1}\vert \leq k_2^{-}+\epsilon_2$.
Note that we also require $\epsilon_1 \leq 1 - k_1^-$ for the square root term to be real-valued. 
Since $F_{\mu}$ is a composition of $C^1$ functions and is thus $C^1$ on $\mathcal{A}^-_{\epsilon}$, by Brouwer's fixed-point theorem \cite[Theorem 52]{Pugh}, the bounding steps above imply the existence of a solution in the set $\mathcal{A}^-_{\epsilon}$, stated below.

\begin{proposition}\label{prop:2bus-fmu-invariance-sol-existence}
    If there exists an $\epsilon$ such that the inequalities in \eqref{eq:2bus-fmu-invariance-eps} hold, then the set $\mathcal{A}_{\epsilon}^-$ is $F_{\mu}$-invariant and 
    the two-bus system possesses a power flow solution in the set $\mathcal{A}_{\epsilon}^-$.
\end{proposition}

As the inequalities \eqref{eq:2bus-fmu-invariance-eps} do not appear to admit straightforward analytical solutions, we will proceed via continuity and argue that  \eqref{eq:2bus-fmu-invariance-eps} are feasible in $\epsilon$ for sufficiently small $\mu$. Of course, any $\epsilon$ that satisfies \eqref{eq:2bus-fmu-invariance-eps} at the boundary 
trivially satisfies the inequalities  themselves; we proceed by focusing on \eqref{eq:2bus-fmu-invariance-eps} with equality sign. Rearranging \eqref{eq:2bus-fmu-invariance-eps} and  moving all terms to one side, we define the mapping $E : \mathcal{D}_{\epsilon} \times \mathbb{R}^4 \goesto \mathbb{R}^2$, where  $\mathcal{D}_{\epsilon} \subset [0, 1-k_1^-) \times [0, 1-k_2^-) \subset \mathbb{R}^2$ is an open set and $E = (E_1, E_2)$ is defined by
\[
\begin{aligned}
    E_1(\epsilon, \mu) &=    \dfrac{-k_{\mu}\gamma_P}{1-k_2^{-}-\epsilon_2} + \dfrac{\gamma_P}{1-k_2^{-}} + \rho\left(1+k_2^{-}+\epsilon_2\right)  \\ &\phantomrel{=}{} + \tilde{\rho}  
        - \epsilon_1 = 0,
    \\ 
    E_2(\epsilon, \mu) &=
        \dfrac{-k_{\mu}\gamma_Q}{1-k_2^{-}-\epsilon_2}
        + \left(1 - k_2^{-} \right)
        + \tilde{\rho} \left(k_1^{-}+\epsilon_1\right)
    \\ &\phantomrel{=}{} - \sqrt{1 - \left(k_1^{-}+\epsilon_1\right)^2}
        -\epsilon_2 = 0.
\end{aligned}
\]
That is, given a $\mu$, an $\epsilon$ that satisfies $E(\epsilon, \mu) = \zero$ satisfies \eqref{eq:2bus-fmu-invariance-eps} with equality sign. Using straightforward algebra and writing $k_1^-, k_2^-$ in terms of the loading margins $\gamma_P, \gamma_Q$, we can verify that when $\mu = \zero$,  we can simply choose $\epsilon  = \zero$ to satisfy $E(\epsilon, \zero) = \zero$ \cite[Proposition 5.1]{Thesis}.  
Using this fact, we can certify that a general local solution to $E(\epsilon, \mu) = \zero$ exists when $\mu \neq \zero$, as summarized below.


\begin{proposition}\label{prop:2bus-fmu-local-sol-ift}
    There exists a nonempty open subset,
    $\mathcal{U} = \mathcal{U}_\epsilon \times \mathcal{U}_{\mu} \subset \mathcal{D}_\epsilon \times \mathbb{R}^4$ and a unique $C^1$ function $f : \mathcal{U}_{\mu} \goesto \mathbb{R}^2$, such that $(\zero_2, \zero_4) \in \mathcal{U}$, $f(\zero_4) = \zero_2$ and $E(f(\mu), \mu) = \zero_2$ for all $\mu \in \mathcal{U}_{\mu}$. That is, for each sufficiently small $\mu$, there exists a unique $\epsilon = f(\mu)$ satisfying  \eqref{eq:2bus-fmu-invariance-eps} with equality sign; consequently, the set $\mathcal{A}_{\epsilon}^-$ contains a power flow solution of the two-bus system.
\end{proposition}


The proof relies on the implicit function theorem \cite{ift}, which we can apply since $E$ is a composition of $C^1$ functions and is thus $C^1$. The existence of a solution then follows from Proposition \ref{prop:2bus-fmu-invariance-sol-existence}. The detailed calculations can be found in \cite[Section 5.3.1]{Thesis}. 
 
Similar to the result in Theorem \ref{thm:2bus-f0-contraction}, the proposition below states that $F_{\mu}$ is further a contraction on $\mathcal{A}^-_{\epsilon}$, so the FPPF algorithm \eqref{eq:2bus-fmu} will converge linearly to the unique solution from any initial condition in $\mathcal{A}^{-}_{\epsilon}$.

\begin{proposition}
    For each sufficiently small $\mu$, there exists $\epsilon$ such that $\mathcal{A}^-_{\epsilon}$ is $F_{\mu}$-invariant and $F_{\mu}$ is a contraction on $\mathcal{A}^-_{\epsilon}$ in the $\ell_{\infty}$ norm.
\end{proposition} 

The proof relies on the fact that $F_{\mu}$ is a $C^1$ function, so we can compute the $\ell_{\infty}$ norm of its Jacobian matrix evaluated at any $\xi_k \in \mathcal{A}_{\epsilon}^-$. Thus, by Theorem \ref{thm:2bus-f0-contraction}, we can argue by continuity that a sufficiently small $\mu$ implies that $F_{\mu}$ is a contraction on $\mathcal{A}^-_{\epsilon}$ in the $\ell_{\infty}$ norm. See \cite[Section 5.3.2]{Thesis} for the detailed derivations. In sum, the above results show that the extended FPPF inherits the convergence properties of the original FPPF \cite{FPPF1}, at least for sufficiently small lossess and transformer tap ratios. Improvement and extension of this analysis to multi-bus systems is a topic of ongoing work.

\section{Numerical Tests}\label{sec:num}

We now illustrate the effectiveness of Algorithm \ref{alg:1} using a selection of \textsc{Matpower} test cases, and compare its behaviour to that of the conventional NR and FDLF methods. Due to space limitations we focus on only two aspects of algorithm performance: (i) iterations required for convergence, and (ii) sensitivity to the initial bus voltage values. For convergence criteria, we set the maximum iteration count to be 100, and the power balance mismatch tolerance to be $10^{-8}$ p.u. for all three algorithms. We use the default single-slack bus model in these test cases since NR and FDLF are not implemented to accommodate the DSB model in \cite{MATPOWER-software}.

\begin{remark}[$\boldsymbol{R/X}$ \textbf{Ratios}]
During testing, it was observed that Algorithm \ref{alg:1} can fail in test cases with unrealistically high branch $R/X$ ratios, which do indeed occur for a small number of branches in certain \textsc{Matpower} test cases. The FDLF and \textemdash{} to a lesser extent \textemdash{} NR algorithms also struggle in cases with high R/X ratios. As practical transmission networks typically have low branch $R/X$ ratios\footnote{The mean $R/X$ ratios for the systems in Table \ref{tab:convergence-iterations} are approximately $0.11, 0.37, 0.14, 0.25, 0.19, 0.15, 0.15$, and $0.19$.}, in the tests that follow we cap all branch $R/X$ ratios at $0.8$, which involves modifying less than 1\% of the branches in each case considered. This modification ensures that the considered cases are convergent for all algorithms. \hfill \oprocend
\end{remark}

\subsection{Iterations Required for Convergence}
\label{Sec:SimIterations}

{
\renewcommand{\arraystretch}{1.0}
\begin{table}[t]
    \centering
    \caption{Iterations required to converge} 
    \vspace{-.5em}
    \begin{tabular}{lrrrrrr}
        \toprule
        & \multicolumn{3}{c}{Base loading} 
        & \multicolumn{3}{c}{High loading}
        \\
        \cmidrule(lr){2-4}
        \cmidrule(lr){5-7}
        Test case & NR & FDLF & FPPF &
        NR & FDLF & FPPF 
        \\ 
        \midrule 
        9 bus system &  4 & 6 & 8 &  5 & 29 & 22 \\
        30 bus system & 3 & 11 & 18 & 6 & 28 & 22  \\
        PEGASE 89 & 4 & 9 & 10 & 6 & 26 & 23 \\
        118 bus system & 4 & 11 & 11 & 6 & 33 & 25 \\
        300 bus system & 5 & 15 & 33 & 6 & 33 & 33 \\
        PEGASE 1354 & 5 & 11 & 42 & 5 & 25 & 42 \\
        PEGASE 2869 & 5 & 11 & 42 & 6 & 29 & 42 \\
        PEGASE 9241 &  6 & 17 & 46 & 6 & 23 & 47 \\
        \bottomrule
    \end{tabular}
    \label{tab:convergence-iterations}
\end{table}
} 

\begin{figure}[t]
    \centering
    \includegraphics{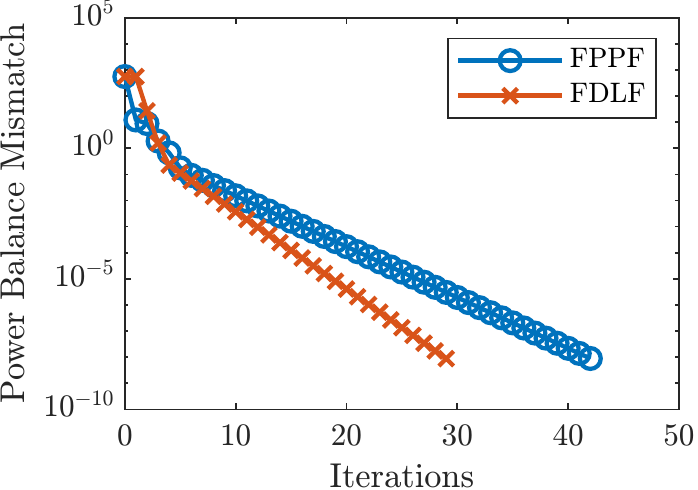}
    \caption{Mismatch tracjectories of FPPF and FDLF on the PEGASE 2869 system with high loading.}
    \label{fig:lin-convergence} 
\end{figure}

Here, we compare the number of iterations each algorithm required to converge using the ``flat-start'' initial condition ($V_L = \one_n$ and $\theta = \zero_n$). 
We present the simulation results based on two loading scenarios: (i) base loading, which is the default values on the test systems, and (ii) high loading, which is computed by continuation power flow (CPF). For the latter scenario, the base power generation and demand are set to be $90\%$
of the way to the power flow insolvability boundary, yielding highly stressed test systems. Table \ref{tab:convergence-iterations} shows the number of iterations each algorithm takes to converge.
When NR converges successfully, it consistently outperforms both FDLF and FPPF due to its quadratic convergence rate. The iterations required by FDLF and FPPF  are comparable, though FDLF is more susceptible to changes in loading level on large systems (300 buses or more). However, both FF and FLDF exhibit linear convergence rates, as shown in Figure \ref{fig:lin-convergence}. 

\subsection{Sensitivity to Initialization}
\label{Sec:SimInitialization}

Next, we test each algorithm's sensitivity to bus voltage initialization. For a constant $\delta \in (0, 1)$,
we generate $1000$ random samples of initial voltage magnitudes $V_{L, \mathrm{init}}^{[k]} \inR^{n}$, where the superscript $[k]$ represents the $k$-th sample. Each element of $V_{L, \mathrm{init}}^{[k]}$ is sampled from a uniform distribution on the interval $[1-\delta, 1+\delta]$, and we set the initial load bus voltage magnitude $V_L^{0}  = V_{L,\mathrm{init}}^{[k]}$, for $k = 1, ..., 1000$ while keeping $\theta = \zero_{n+m}$. 

We first compute the known high voltage solution using NR with flat-start voltages, then compare this solution against the ones returned by NR, FDLF and FPPF using the random voltage initializations. If the solution returned does not match the known solution up to a small tolerance, or if the algorithm fails to converge, then the sample is marked as unsuccessful, otherwise it is successful. 
Table \ref{tab:init-118} demonstrates each algorithm's success rate for different $\delta$ on the 118 bus system.   

{
\renewcommand{\arraystretch}{1.0}
\begin{table}
    \centering
    \caption{Algorithm success rate (\%), 118 bus system}
    \vspace{-.5em}
    \begin{tabular}{lrrrrrr}
        \toprule
        & \multicolumn{3}{c}{Base loading} 
        & \multicolumn{3}{c}{High loading}
        \\
        \cmidrule(lr){2-4}
        \cmidrule(lr){5-7}
        $\delta$ & NR & FDLF & FPPF &
        NR & FDLF & FPPF 
        \\ 
        \midrule 
        0.1 & 100.0 & 100.0 & 100.0 & 100.0 & 100.0 & 100.0\\
        0.2 & 98.8 & 100.0 & 100.0 & 98.8 & 100.0 & 100.0 \\
        0.3 & 65.6 & 100.0 & 100.0 & 66.7 & 100.0 & 100.0 \\
        0.4 & 6.9 & 100.0 & 100.0 & 7.7 & 100.0 & 100.0 \\
        0.5 & 0.0 & 100.0 & 100.0 & 0.0 & 100.0 & 100.0 \\       
        0.9 & 0.0 & 100.0 & 100.0 & 0.0 & 98.7 & 100.0 \\
        0.95 & 0.0 & 100.0 & 100.0 & 0.0 & 89.0 & 98.9 \\
        \bottomrule
    \end{tabular}
    \label{tab:init-118}
\end{table}
}
Evidently, NR is the least robust against the random voltage magnitude initialization since its success rate drastically decreases as $\delta$ increases. This observation matches the well-known fact that the convergence of NR is extremely sensitive to the initial condition selection \cite{nr}. FDLF performs almost as well as FPPF until $\delta$ gets close to $1$ in the high loading scenario, where a small number of samples fail to converge. 

When the FPPF converges, it always converges to the known high voltage solution, and is the most robust out of all three algorithms under this random initialization scheme. This may be a valuable feature in power flow problems where there is considerable uncertainty about the location of the solution. Alternatively, the FPPF may be valuable as a warm-start tool for a NR-based solver. Curiously, when the FPPF algorithm fails, it is because during the iterations, the implicit constraint that $\Vert \psi \Vert_{\infty} \leq 1$ in \eqref{eq:psi-solution-full} is violated; this constraint keeps $\psi$ real-valued. A rigorous procedure to ensure that this constraint remains satisfied during iterations is a subject of ongoing work. 

\section{Conclusion}

We have derived and tested a new algorithm for the AC power flow problem by extending the lossless FPPF algorithm of \cite{FPPF1} to accommodate network loss, phase-shifting transformers, and the distributed slack bus model. As a first step in the theoretical analysis of the algorithm, we studied it on the two-bus system and presented sufficient conditions for the algorithm to converge to the desired solution. We also tested the numerical performance of the proposed FPPF algorithm on standard small- and large-scale test cases. Avenues of future work include improving the algorithm's robustness against branches with high $R/X$ ratios (potentially by modifying the fixed-point reformulation) and extending the convergence conditions of the proposed FPPF algorithm on the two-bus system to general systems. 

\printbibliography
\end{document}